# Anisotropic magnetic properties of the triangular plane lattice material TmMgGaO$_4$

F. Alex Cevallos, Karoline Stolze, Tai Kong, and R.J. Cava

Department of Chemistry, Princeton University, Princeton NJ 08544 USA

*Corresponding authors: fac2@princeton.edu (F. Alex Cevallos), rcava@princeton.edu (R.J. Cava)

**Abstract**

The crystal growth, structure, and basic magnetic properties of TmMgGaO$_4$ are reported. The Tm ions are located in a planar triangular lattice consisting of distorted TmO$_6$ octahedra, while the Mg and Ga atoms randomly occupy intermediary bilayers of M-O triangular bipyramids. The Tm ions are positionally disordered. The material displays an antiferromagnetic Curie Weiss theta of ~ -20 -25 K, with no clear ordering visible in the magnetic susceptibility down to 1.8 K; the structure and magnetic properties suggest that ordering of the magnetic moments is frustrated by both structural disorder and the triangular magnetic motif. Single crystal magnetization measurements indicate that the magnetic properties are highly anisotropic, with large moments measured perpendicular to the triangular planes. At 2 K, a broad step-like feature is seen in the field-dependent magnetization perpendicular to the plane on applied field near 2 Tesla.

**Introduction**

Geometrically frustrated magnetism has been an actively studied property of materials since at least the 1970s[1-2]. In a geometrically frustrated system, the geometry of the crystal lattice inhibits the long-range ordering of the magnetic moments; the simplest examples are systems with antiferromagnetic nearest neighbor interactions on a triangular lattice. As there is no simple way to order the spins in such a system, the magnetic ordering transition will typically be below the Weiss temperature. Geometrically frustrated systems can be used to study a variety of unusual ground state conditions that are difficult to achieve in more conventional materials[1-7]. The influence of structural disorder, especially in geometrically frustrated rare-earth based systems, is expected to be significant[4].

The $YbFe_2O_4$ structure type has been observed to demonstrate a wide variety of electronic and magnetic properties, including geometric magnetic frustration, spin-glass behavior[7-11], charge density waves[12-13], and ferroelectricity[14-16]. This structure crystallizes in the *R-3m* space group, and is defined by triangular planes of metal-oxygen octahedra separated by a bilayer of metal-oxygen triangular bipyramids[17]. Recently, the $YbFe_2O_4$-type compound $YbMgGaO_4$ has drawn interest as a potential candidate for exhibiting quantum spin liquid (QSL) behavior[18-21]. $YbMgGaO_4$ places the magnetic $Yb^{3+}$ ions on the triangular planes, with non-magnetic $Mg^{2+}$ and $Ga^{3+}$ ions randomly mixed in the triangular bipyramidal sites (but with no mixing between magnetic and nonmagnetic ions). The Mg-Ga mixing, off the magnetic plane, causes a considerable variation in local magnetic interactions that is reflected in the compound's properties[20, 22-23]. Here we report the structure and elementary magnetic properties of the isostructural and closely related compound $TmMgGaO_4$, in which similarly complex magnetic behavior is possible. Our results suggest that further, more detailed study may be of future interest.

**Experimental**

*Synthesis*

Polycrystalline samples of $TmMgGaO_4$ were synthesized via solid state reaction. Stoichiometric quantities of $Tm_2O_3$ (99.99%, Alfa Aesar), $Ga_2O_3$ (99.999%, Alfa Aesar) and MgO (99.95%, Alfa Aesar) were ground with an agate mortar and pestle. Pellets of the starting composition were heated in ceramic crucibles in air at 1450°C for four days, with intermediate grinding. A small amount of impurity, believed to be $Tm_3Ga_5O_{12}$, was observed in some samples and attributed to an insufficient starting quantity of MgO due to partial hydration of the starting material. The addition of MgO to the impure sample at the 5% level, followed by reheating at 1450°C for 24 hours resulted in the reduction or elimination of this impurity in all cases.

Single crystals in the mm size range were grown by the floating zone method. Single phase polycrystalline powder was loaded into rubber tubes and hydrostatically compressed at 40 MPa, resulting in polycrystalline rods typically measuring 6 mm in diameter and 5-8 cm in length. The rods were then sintered at 1450°C for 3 hours in air before being transferred to a four-mirror optical floating zone (FZ) furnace (Crystal Systems, Inc. Model No. FZ-T-10000-HVP-II-P) with 4 x 1000 W lamps. The rods were then further sintered in the FZ furnace at 65.0% output power, rotating at 15 RPM, with a rate of travel of 1.5 mm/hr. Crystal growth from the feed rods was performed at 68.0% output power, with rods rotating in opposite directions at 20 RPM and an upwards travel rate of 0.5 mm/hr. Growth of faceted millimeter-scale crystals of $TmMgGaO_4$ was observed within a few mm of growth initiation, but large single crystals were not obtained for growths up to 5 cm in length, even when using previous growths as a seed.

*Characterization*

Room-temperature powder X-ray diffraction (PXRD) measurements were performed with a Bruker D8 Advance Eco diffractometer with Cu $K\alpha$ radiation ($\lambda = 1.5418$ Å) and a LynxEye-XE detector. Phase identification was performed using the Bruker EVA program. Powder Rietveld refinements were performed using Fullprof Suite. Magnetic measurements were taken using a Quantum Design Physical Property Measurement System (PPMS) Dynacool with a vibrating sample mount. DC magnetic susceptibility, defined as the measured magnetization *M* divided by the applied magnetic field, was measured between 1.8 K and 300 K in an applied field of 1000 Oe, and the resulting values were divided by the number of moles of $Tm^{3+}$ present to obtain the magnetization values per formula unit. Field-dependent magnetization was measured at 2 K. Powder samples for magnetic characterization were obtained by grinding small single crystals. Anisotropic magnetization measurements were taken on a 1 x 0.2 mm plate-like single crystal (large face perpendicular to the c-axis), either mounted on a silica sample holder with GE varnish or placed in a plastic sample holder, and oriented along the axis of study.

**Results and Discussion**

*Structure*

TmMgGaO$_4$, as has been previously reported[24], is isostructural to YbFe$_2$O$_4$, and therefore to a large family of related isostructural $Ln^{+2}M^{2+}M^{3+}O_4$ compounds[8, 25-26]. In agreement with previous reports, our colorless, transparent TmMgGaO$_4$ crystallizes in the space group *R-3m* (166), with lattice parameters of *a* = 3.4195(3) Å and *c* = 25.1231(1) Å. The structure is composed of triangular layers of distorted TmO$_6$ octahedra, separated by bilayers of mixed occupancy Mg- and Ga-O triangular bipyramids (Figure 2). The Tm atoms sit displaced, slightly off the ideal [0 0 0] position along *c*, with z = 0.0078(2). This distortion has been observed in other materials in this family and is attributed to the random distribution of non-lanthanide metals in the neighboring bilayers[8]. As the Tm atom can only occupy one of the resulting displaced sites, they are both determined to have 1/2 occupancy (the sites are too close together to be simultaneously occupied). Although the lattice parameters and overall structure are in good agreement with previously published results[24-25], the disordered slightly off-ideal location of the Tm$^{3+}$ ion did not appear to be noted in the only previous study reporting refined atomic positions[25]. However, this positional disorder of the rare earth atom has been observed in a variety of related systems, including LuCuGaO$_4$, LuCoGaO$_4$, YbCuGaO$_4$, LuCuFeO$_4$, LuZnFeO$_4$[8], and YbMgGaO$_4$[21], suggesting that TmMgGaO$_4$ behaves similarly. Our single-crystal XRD refinement of the structure (see Supplementary Information) showed that the statistic positional disorder of the Tm positions can also be described by an approximately equivalent model wherein the Tm atom sits directly on the high-symmetric Wyckoff site *3a* [0 0 0] site with a very large anisotropic displacement parameter along the *c*-axis. The results of this refinement can be seen in the supplementary information, along with a map of the measured atomic scattering density showing the distribution of thulium positions in a section of the unit cell, derived from the observed structure factors, $F_{obs}$.

*Magnetism*

The temperature-dependent magnetic susceptibility of a powder sample of TmMgGaO$_4$ (sample produced by grinding a collection of small single crystals) is shown in Figure 3. The susceptibility was fit to the Curie-Weiss law $\chi - \chi_0 = C / (T - \theta_W)$, where $\chi$ is the magnetic susceptibility, $\chi_0$ is a temperature-independent contribution, $C$ is the Curie constant, and $\theta_W$ is the Weiss temperature. The inverse susceptibility, $1/(\chi - \chi_0)$, was found to be almost linear for a $\chi_0$ value of 0.0010 emu mol$^{-1}$. The $\chi_0$ value for the polycrystalline sample is consistent with the single crystal susceptibility obtained for fields in the plane of the triangular lattice, as described below. The inverse susceptibility was fit at high and low temperature. At high temperature (150 K – 290 K), $C$ was found to be 7.26 and $\theta_W$ was found to be -25.6 K. The effective magnetic moment per ion, $\mu_{eff}$, was determined by the relationship $\mu_{eff} = \sqrt{8C}$, yielding a moment of 7.62 $\mu_B$/Tm, in good agreement with the ideal value for a free Tm$^{3+}$ ion, 7.57 $\mu_B$. At low temperatures, a linear fit yielded values of $C$ = 6.68, $\theta_W$ = -19.7, and an effective magnetic moment of 7.31 $\mu_B$/Tm. The negative Weiss temperatures determined by these fits suggest that TmMgGaO$_4$ has dominantly antiferromagnetic coupling, but no magnetic ordering is observed down to 1.8 K. At the lowest temperatures studied here (1.8-10 K) there is a deviation from the paramagnetic behavior predicted by the higher temperature Curie Weiss fits (Figure 3, inset). TmMgGaO$_4$ therefore exhibits a frustration index $f$ ($\theta_W$ /$T_M$) of 10 or more, suggesting that it is a strongly frustrated magnet[4].

A magnetization vs. applied field measurement was taken on the same powder sample at 2 K. The material exhibited a nonlinear magnetization response with an apparent saturation of approximately 5 $\mu_B$/Tm by applied fields of 8 Tesla. Repeating this measurement on a small single crystal of TmMgGaO$_4$ perpendicular to the triangular plane (i.e. parallel to the *c*-axis) resulted in a much higher saturation magnetization, of approximately 7 $\mu_B$/Tm (Figure 4). Both the powder and single-crystal magnetizations demonstrate a slope anomaly with an onset field of approximately $\mu_0 H \sim$ 2 Tesla. The origin of this anomaly is currently unknown, but its appearance is reminiscent of metamagnetism and its presence is consistent across all samples. In contrast, magnetization vs. applied field measurements conducted parallel to the triangular planes (i.e. perpendicular to *c*) yielded very different results: a much smaller magnetization with a very slight curvature and no apparent saturation. From this it can be safely concluded that TmMgGaO$_4$ has a high degree of magnetic anisotropy, which we attribute to crystal electric field (CEF) effects originating from its layered crystal structure and distorted TmO$_6$ octahedra. This distortion of the octahedra results from the previously-described off-site nature of the central Tm atoms, which is itself attributed to the local distortions caused by disorder on the Mg/Ga sites. We note that strong CEF effects have been found in the isostructural compound YbMgGaO$_4$[20], lending credence to our hypothesis concerning the origin of the magnetic anisotropy.

In order to observe this anisotropy more clearly, magnetization vs temperature measurements were taken on a single crystal (pictured in Figure 1), both parallel and perpendicular to the triangular layers. As can be seen in Figure 5, the anisotropy is clearly displayed in this measurement as well. Measurements parallel to the perpendicular to the triangular planes (i.e. along the *c*-axis) yielded a similar curve to the powder sample, but much larger values of molar magnetic susceptibility. Measurements parallel to the planes show a much weaker signal, with a broad region spanning from 50 K to 300 K that appears to be nearly temperature-independent. The purple curve in Figure 5 is a weighted average of the temperature-dependent susceptibilities of the single crystal (1/3 of the value parallel to the *c*-axis plus 2/3 of the value perpendicular to the c-axis). This calculated value overlaps almost exactly with the measured susceptibility of the powder. In the right panel of Figure 5, the inverse is shown of all four susceptibilities. The weighted average of the anisotropic susceptibilities in this case is more linear than the unmodified inverse susceptibility of the powder, but when the term $\chi_0$ (which we attribute to the container for the powder sample) is applied to the powder measurement, the two once again overlap almost exactly. The observation that the magnetic moment as measured in a polycrystalline powder is effectively a simple average of the moments along the spatial directions is consistent with the hypothesis that this anisotropy arises as a result of CEF effects within the crystal lattice[27].

**Conclusion**

Single crystals of TmMgGaO$_4$ have been synthesized in an optical floating zone furnace, and the structure has been refined. The resulting values for lattice parameters and atomic positions are in good agreement with previously published results for this and other isostructural compounds. Magnetization measurements were performed, finding the material to display dominantly antiferromagnetic interactions, with no ordering above 1.8 K. Thus the magnetic characterization suggests a magnetic frustration index of 10 or more. Magnetization measurements on a single crystal show that the magnetic properties are highly anisotropic, and suggest that this anisotropy may arise from CEF effects. Further characterization of the low temperature properties of this material is warranted; the uncommon interplay of an isolated, geometrically-frustrated magnetic 2D lattice, and the subtle structural distortions induced on the magnetic interactions due to the off-plane disorder of non-magnetic ions may give rise to unusual electronic or magnetic ground states, and similar to the disordered transition metal pyrochlores, may provide interesting system for testing the effect of disorder in geometrically frustrated magnets[28-33]. Further efforts to grow larger high-quality single crystals for complimentary measurement purposes, such as neutron scattering, may also be of interest.

**Acknowledgments**

This research was supported by the US Department of Energy, Division of Basic Energy Sciences, Grant No. DE-FG02-08ER46544, and was performed under the auspices of the Institute for Quantum Matter.

**Table 1.** Refined crystal structure of TmGaMgO$_4$ at ambient temperature, space group *R-3m* (no. 166), unit cell parameters *a* = 3.4195(3) Å and *c* = 25.1231(12) Å, from the ambient temperature powder X-ray diffraction data. $R_f$ factor = 17.1, $R_p$ = 10.1, $R_{wp}$ = 14.7, $\chi^2$ = 13.7

| Atom | Wyckoff site | *x* | *y* | *z* | Occ. |
|---|---|---|---|---|---|
| Tm | 6*c* | 0 | 0 | 0.0078(2) | 0.5 |
| Mg | 6*c* | 0 | 0 | 0.2147(2) | 0.5 |
| Ga | 6*c* | 0 | 0 | 0.2147(2) | 0.5 |
| O1 | 6*c* | 0 | 0 | 0.2836(4) | 1 |
| O2 | 6*c* | 0 | 0 | 0.1368(4) | 1 |

**Figure Captions**

**Figure 1.** Powder Rietveld refinement of TmMgGaO$_4$. Inset: A small, colorless transparent single crystal of TmMgGaO$_4$ on a 1-mm grid.

**Figure 2.** The crystal structure of TmMgGaO$_4$ showing the coordination polyhedra. TmO$_6$ octahedra are blue, GaO$_5$ or MgO$_5$ triangular bipyramids are orange. On the right is the triangular magnetic lattice formed by the Tm layers, as viewed along the *c*-axis. In the bottom right is a closer look at the distorted nature of the TmO$_6$ octahedra.

**Figure 3.** Left panel: The temperature-dependent DC magnetic susceptibility and reciprocal susceptibility of a polycrystalline powder of TmMgGaO$_4$ in an applied field of 1000 Oe. Curie-Weiss fits are shown in black. Right panel: The field-dependent magnetization at 2 K. Inset: Magnified view of the low-temperature region of the inverse susceptibility.

**Figure 4.** Magnetization vs applied field measurements on a single crystal of TmMgGaO4 parallel to the c-axis, perpendicular to the *c*-axis, and on a powder sample of the same material. Inset: Derivative of magnetization parallel to the *c*-axis with respect to applied field.

**Figure 5.** Left panel: The temperature-dependent DC susceptibility of a single crystal of TmMgGaO$_4$ as measured parallel to the *c*-axis (red) and perpendicular to the *c*-axis (blue). In purple is a weighted average of the directional susceptibilities. The measured susceptibility of the polycrystalline powder is superimposed in green. Right panel: The reciprocal susceptibilities of the single crystal and polycrystalline powder.

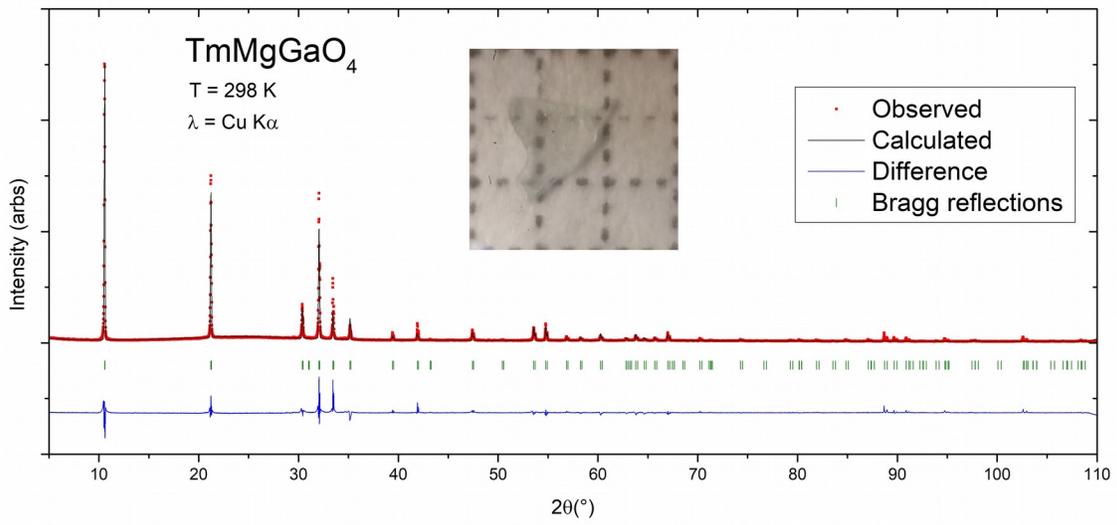

**Figure 1.**

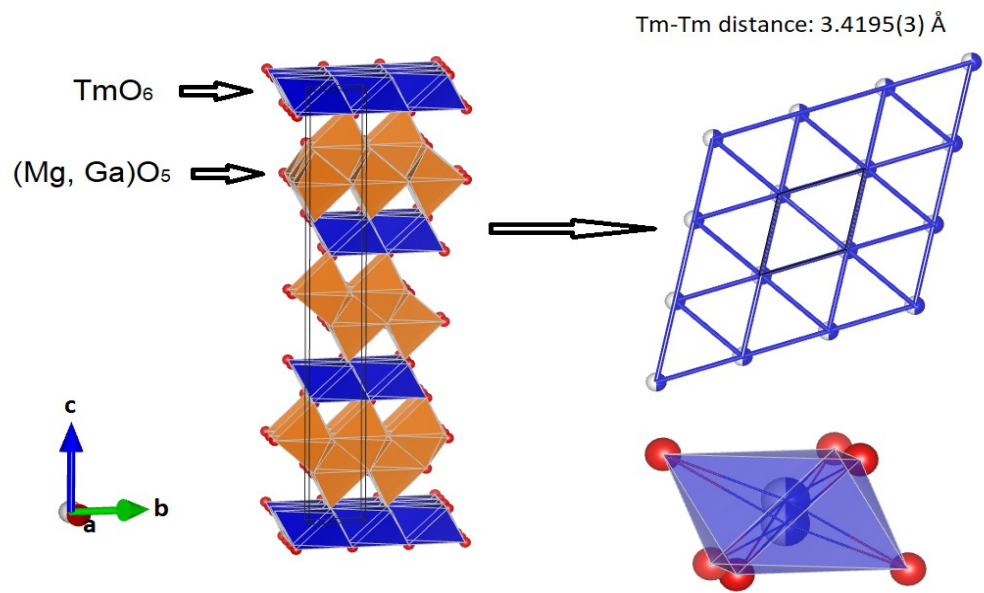

**Figure 2.**

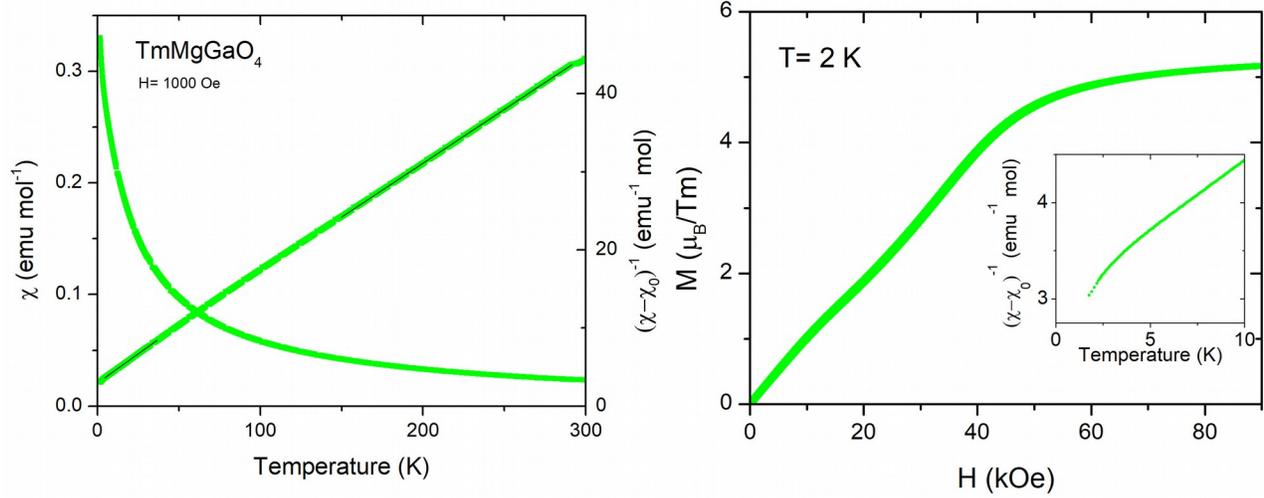

**Figure 3**

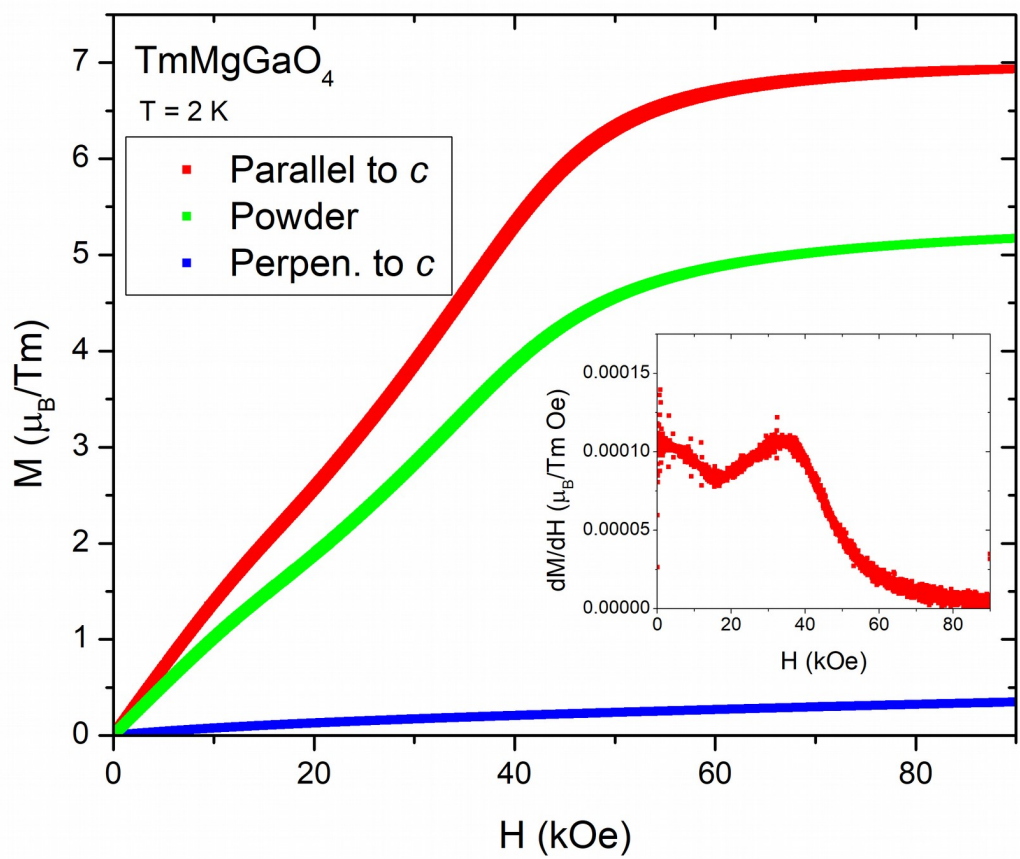

**Figure 4.**

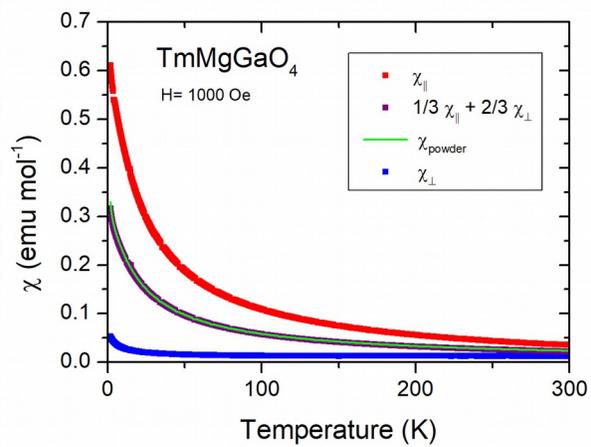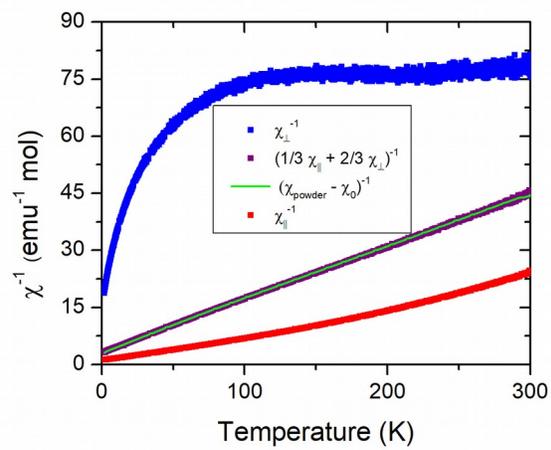

**Figure 5**

**Supplementary Information**

**Experimental**

The crystal structure of TmMgGaO$_4$ was further determined by single-crystal X-ray diffraction (SXRD), especially to analyze the positional disorder of Tm and to evaluate the Ga/Mg split position. The SXRD data was collected at 296 K with a Kappa APEX DUO diffractometer equipped with a CCD detector (Bruker) using graphite-monochromatized Mo-$K\alpha$ radiation ($\lambda$ = 0.71073 Å). The raw data was corrected for background, polarization, and the Lorentz factor using APEX2 software[i], and multi-scan absorption correction was applied.[ii] The structure was solved with the charge flipping method[iii] and subsequent difference Fourier analyses with Jana2006. [iv, v, vi] Structure refinement against $F_o^2$ was performed with Shelxl-2017/1.[vii, viii]

The initial structure model with anisotropic refined thermal parameters for Tm revealed a broad electron distribution on the Tm1 position (Wyckoff site 3$a$, [0 0 0]) along the $z$-axis. Thus, a new site (Wyckoff site 6$c$) was introduced to describe the statistic positional disorder of the rare earth atom, which could be refined to [0, 0, 0.00528(3)] using an isotropic thermal parameter and 50 % occupancy for Tm. The new coordinates were taken as start position for the powder Rietveld refinement. As the refinement of the anisotropic thermal parameters $U_{ij}$[Tm1] results in a shift of the coordinates back to [0 0 0], the high symmetric site 3$a$ with a large displacement parameter along $z$ was used to describe the positional disorder of Tm in the final single-crystal structure model. Ga and Mg share a split position with 50% occupancy each. Due to correlation of parameter values the displacement parameters and coordinates for Ga and Mg were restricted be the same.

The examination of the Fourier ($F_{obs}$) map (Figure S1) revealed significant elongated electron density maxima on the Tm1 position along $z$, clearly indicating the positional disorder of the rare earth metal, and a less distinct but still prolate electron distribution on the Ga1/Mg1 split position along $z$ as well, which also points towards a positional disorder in addition to the mixed occupancy. Crystallographic data are summarized in Table S1, final atomic parameters are listed in Table S2 and S3.

**Table S1.** Crystallographic data and details of the structure determination of TmMgGaO$_4$ derived from single-crystal experiments measured at 296(1) K.

| Sum Formula | TmMgGaO$_4$ |
| --- | --- |
| Formula weight / (g · mol$^{-1}$) | 326.96 |
| Crystal System | trigonal |
| Space group | $R\bar{3}m$ (no. 166) |
| Formula units per cell, $Z$ | 3 |
| Lattice parameter $a$ / Å | 3.4250(6) |
| $c$ / Å | 25.169(4) |
| Cell volume / (Å$^3$) | 255.7(1) |
| Calculated density / (g · cm$^{-3}$) | 6.370 |
| Radiation | (Mo-$K\alpha$) $\lambda$ = 0.71073 Å |
| Data range | $2\theta \leq 82.23°$ <br> $-6 \leq h \leq 6$ <br> $-6 \leq k \leq 6$ <br> $-45 \leq l \leq 44$ |
| Absorption coefficient / mm$^{-1}$ | 33.79 |
| Measured reflections | 3119 |
| Independent reflections | 262 |
| Reflections with $I > 2\sigma(I)$ | 249 |
| $R$(int) | 0.024 |
| $R(F_o^2)$ | 0.012 |
| No. of parameters | 10 |
| $R_1$(obs) | 0.010 |
| $R_1$(all $F_o$) | 0.012 |
| $wR_2$(all $F_o$) | 0.022 |
| Residual electron density / (e · Å$^{-3}$) | 0.67 to $-0.74$ |

**Table S2.** Wyckoff positions, coordinates, occupancies, equivalent and isotropic displacement parameters respectively for TmMgGaO$_4$ single-crystal measured at 296(1) K. The coordinates of G1 and Mg1 were equalised; $U_{eq}$ is one third of the trace of the orthogonalized $U_{ij}$ tensor.

| Atom | Wyck. Site | $x$ | $y$ | $z$ | Occupancy | $U_{eq}/U_{iso}$ |
| --- | --- | --- | --- | --- | --- | --- |
| Tm1 | 3$a$ | 0 | 0 | 0 | 1 | 0.01154(5) |
| Ga1 | 6$c$ | 0 | 0 | 0.21437(2) | 0.5 | 0.00580(6) |
| Mg1 | 6$c$ | 0 | 0 | 0.21437(2) | 0.5 | 0.00580(6) |
| O1 | 6c | 0 | 0 | 0.29101(7) | 1 | 0.0085(3) |
| O2 | 6c | 0 | 0 | 0.12881(8) | 1 | 0.0144(3) |

**Table S3.** Anisotropic displacement parameters for TmMgGaO$_4$ single-crystal measured at 296(1) K. The coefficients $U_{ij}$ (/Å$^2$) of the tensor of the anisotropic temperature factor of atoms are defined by $\exp\{-2\pi^2[U_{11}h^2a^{*2} + \cdots + 2U_{23}klb^*c^*]\}$; $U_{ij}$[Ga1] and $U_{ij}$[Mg1] were equalised.

| Atom | $U_{11}$ | $U_{22}$ | $U_{33}$ | $U_{12}$ |
|---|---|---|---|---|
| Tm1 | 0.00417(4) | 0.00417(4) | 0.02627(9) | 0.00208(2) |
| Ga1/Mg1 | 0.00505(8) | 0.00505(8) | 0.00730(15) | 0.00253(4) |

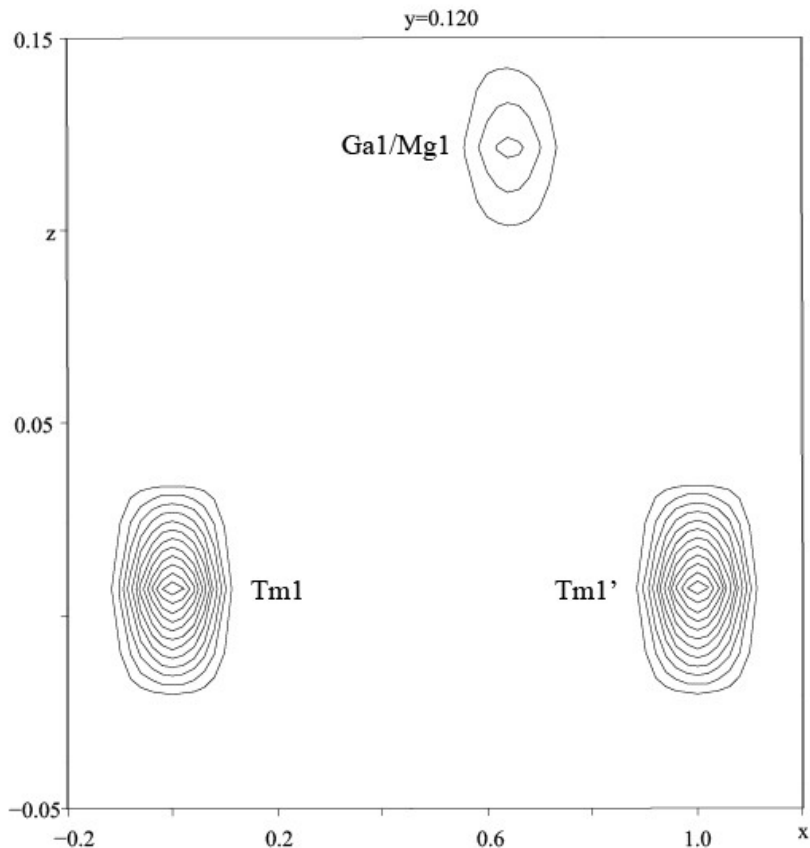

**Figure S1.** Fourier map ($F_{obs}$) for TmMgGaO$_4$ based on room-temperature data calculated in space group $R\bar{3}m$, map summed up between $-0.10 < y < 0.35$, contour lines correspond to 10 e/Å$^3$.